\documentclass[10pt,conference]{IEEEtran}
\IEEEoverridecommandlockouts
\usepackage{cite}
\usepackage{amsmath,amssymb,amsfonts}
\usepackage{algorithmic}
\usepackage{url}
\usepackage{booktabs}
\usepackage{multicol}
\usepackage{graphicx}
\usepackage{textcomp}
\usepackage[table]{xcolor}

\usepackage{hyperref}
\usepackage{balance}
\def\BibTeX{{\rm B\kern-.05em{\sc i\kern-.025em b}\kern-.08em
    T\kern-.1667em\lower.7ex\hbox{E}\kern-.125emX}}

\begin{document}
\pagestyle{plain}

\title{Do Newer Models Produce Better Patches? A Non-Functional Quality Study}


\author{
\IEEEauthorblockN{Xin Sun, Daniel St\aa hl, Kristian Sandahl, Christoph Kessler}
\IEEEauthorblockA{
Department of Computer and Information Science\\ Linköping University, Sweden\\
\{xin.sun, daniel.stahl, kristian.sandahl, christoph.kessler\}@liu.se }}

\maketitle

\begin{abstract}

Repository-level coding benchmarks typically measure progress in model capability by comparing the resolved rates of later and earlier models. However, this focus overlooks whether the non-functional quality of their generated patches has also changed across model generations. This study investigates whether later models produce functionally correct patches with better non-functional characteristics than earlier models on comparable repository-level repair tasks. We conducted two case studies involving four Claude and DeepSeek models on SWE-bench Lite. Using the same SWE-agent configuration, we evaluated the generated patches with CodeQL, CodeScene, CPU time, and peak memory. Our primary analysis compared the models on commonly resolved instances. The static analysis results showed that most CodeQL paired differences were zero and that no CodeQL or CodeScene comparison remained significant after Holm correction. CPU time differences were small and inconsistent across model families, while peak memory usage was slightly higher for the later models under the benchmark test workload, with small absolute differences. Differences in individual CodeQL rules and CodeScene categories varied across model families and did not survive multiple-comparison correction. Overall, later models resolved more instances but showed no consistent improvement in the measured non-functional indicators on tasks solved by both models. Through this study, we hope to encourage a more comprehensive evaluation of models' practical software engineering capabilities in the future.
\end{abstract}


\section{Introduction}

The widespread use of LLMs has significantly accelerated code generation and feature implementation in the software development process~\cite{xia2025agentless}. As LLMs continue to evolve, newer generations of models often provide longer context windows, stronger reasoning capabilities, and better performance on repository-level software engineering tasks~\cite{jimenez2024swebench}. These advances have improved the ability of LLMs to address real-world software issues and to generate functionally correct patches. However, software quality is not limited to functional correctness. It also includes a wide range of non-functional quality attributes, such as maintainability, security, and reliability.

Existing benchmarks for software engineering tasks, such as SWE-bench~\cite{jimenez2024swebench}, primarily evaluate model capability by checking whether the generated patches pass the provided test suites and by reporting the resolved rate. This evaluation provides a clear and practical measure of functional repair success, but it offers only a limited indication of the non-functional quality of the generated patches. As a result, such evaluations may overestimate the practical quality of LLM-generated patches in real-world software engineering settings~\cite{sun2025qualityassurancellmgeneratedcode}. Even when newer models generate more functionally correct patches, their code may still introduce new code smells, security risks, or potential performance regressions, as older models do~\cite{coig2024Aperformance,pearce2022asleep}.

To obtain a more complete understanding of model behavior in software engineering tasks, this study investigates the non-functional quality of patches generated by different LLMs on SWE-bench Lite. The prompt asks the models to repair a functional issue and does not explicitly request improvements in any specific non-functional quality dimension, such as security, reliability, or performance efficiency. We use two static analysis tools to assess changes in selected static quality aspects of the functionally correct patches. We also measure CPU time and peak memory usage during the benchmark tests to characterize the runtime behavior of the patches.

Through this analysis, we examine whether improvements in functional correctness are accompanied by corresponding improvements in patch quality on comparable tasks. We compare an earlier model and a later model in each of two families. The study is guided by the following research questions:

\begin{enumerate}

    \item[\textbf{RQ1.}] Among commonly resolved instances, to what extent do later and earlier models differ in the non-functional quality of their patches?

    \item[\textbf{RQ2.}] What non-functional quality patterns can be observed in patches generated by later and earlier model configurations?

\end{enumerate}

\section{Background}
\subsection{LLMs in Software Engineering}

The application of LLMs in software engineering has expanded from function-level code generation to repository-level tasks such as issue understanding, bug fixing, code modification, and refactoring. These tasks require models to reason over natural language descriptions, project structure, source code, dependencies, test feedback, and tool outputs. Single-prompt interactions are increasingly replaced by agent frameworks, which enable the model to browse repository files, execute commands, edit code, and iteratively revise its solution based on feedback~\cite{yang2024sweagent,zhang2024autocoderover,wang2025openhands,xia2025agentless}. Coding agents such as Claude Code and Codex~\cite{chen2021codex} instantiate this agent-based workflow and can operate over codebases to solve complex tasks. In repository-level tasks, the output of the model is no longer an isolated snippet but a patch that can be applied directly to a real codebase. Patch quality therefore bears directly on long-term maintenance cost, and its evaluation must extend beyond whether the code is functionally correct.

\subsection{Functional Correctness Evaluation}
Repository-level software engineering benchmarks like SWE-bench commonly evaluate functional correctness by testing whether a generated patch can be applied successfully and pass the evaluation harness. This test-based evaluation is easy to reproduce and compare across models, which is why metrics such as resolved rate are widely used to report LLM performance on software engineering tasks.

Early benchmarks such as HumanEval~\cite{chen2021codex} and MBPP~\cite{austin2021mbpp} evaluate function-level code generation using unit tests. Defects4J~\cite{just2014defects4j}, BugsInPy~\cite{widyasari2020bugsinpy}, and QuixBugs~\cite{lin2017quixbugs} extend test-based evaluation to program repair. SWE-bench~\cite{jimenez2024swebench} further constructs tasks from real GitHub issues and their corresponding pull requests, placing models in repository-level repair settings that more closely approximate real software-maintenance activities. Subsequent work has refined or extended this benchmark, including a human-validated subset~\cite{openai2024swebenchverified} and a multilingual extension covering seven additional languages~\cite{zan2025multiswebench}. In this work, we adopt the Lite subset of SWE-bench to maintain comparability with reported leaderboard results; the precise configuration is described in \autoref{sec:method}.

Passing the benchmark tests, however, captures only the behavior covered by those tests. A functionally correct patch may still introduce unnecessary complexity, duplicated logic, fragile assumptions, inadequate exception handling, static quality issues, or increased resource consumption. These concerns are especially important for repository-level tasks, where a generated patch becomes part of an evolving codebase and may affect future maintenance. This makes functional correctness a necessary first step, but not a complete basis for evaluating generated patches.

\subsection{Non-functional Quality Evaluation}
Passing functional tests does not guarantee that generated code has comparable security, maintainability, or performance efficiency~\cite{pearce2022asleep,sun2025qualityassurancellmgeneratedcode}. Fu et al.~\cite{fu2025security} analyzed Copilot-generated code in GitHub projects and reported instances of unsafe API usage, improper input handling, and multiple CWE security weaknesses. This suggests that code satisfying a local functional requirement may still introduce latent security risks. Besides security weaknesses, generated code may also contain duplicated logic, complex control flow, overly specialized conditions, and inconsistencies with the surrounding codebase~\cite{siddiq2022codesmells}. Performance efficiency is another concern. Functional tests usually check whether the program produces the expected behavior, but they do not systematically measure execution time, memory usage, or resource overhead. A repository-level patch may alter execution paths, add extra computation, or increase peak memory usage without causing test failures.

Existing studies have examined non-functional quality in different repair settings. Sun et al.~\cite{sun2025qualityassurancellmgeneratedcode} evaluated functionally correct SWE-bench Lite patches using static analysis and resource measurements and also examined feedback targeting specific quality dimensions. Wadhwa et al.~\cite{wadhwa2024core} and Patcas and Motogna~\cite{patcas2026evaluation} instead evaluated repairs prompted by static-analysis findings.

These studies show that non-functional quality varies across models and repair settings. However, they do not examine whether later models with higher resolved rates also produce better-quality patches on tasks resolved by both models in a provider pair. Comparisons based on each model's full resolved set may instead reflect differences in task composition.

An earlier study~\cite{sun2025qualityassurancellmgeneratedcode} established the evaluation setting used here by assessing the non-functional quality of functionally correct SWE-bench Lite patches. The present study reuses previously generated patches for models shared with that study and evaluates newly generated patches for the additional models under the same functional repair setting. It addresses a different research question by conducting two case studies, each comparing an earlier and a later model from the same provider. The primary analysis focuses on instances resolved by both models and combines static analysis with repeated measurements of CPU time and peak memory. The matched-instance comparisons, resolved-set selection analysis, patch-size sensitivity analysis, and their corresponding results have not been reported previously.

\section{Method}
\label{sec:method}
This section describes our procedure for evaluating the non-functional quality of LLM-generated patches. For each instance, we obtained patches generated with SWE-agent using different model versions. We then evaluated functional correctness using the benchmark harness, retained functionally correct patches for non-functional analysis, and assessed each patch using static analysis tools and the selected runtime and memory measurements. Finally, we conducted statistical analyses to compare the non-functional quality metrics across different models.

\subsection{Experiment Design}
\paragraph{Benchmark}

In this study, we used SWE-bench Lite as the benchmark and SWE-agent as the agent framework. SWE-bench is a benchmark designed to evaluate the performance of LLMs on real-world software engineering tasks. It consists of repository-level issues extracted from GitHub, requiring models to navigate and understand the repository context, coordinate changes across multiple files, and generate patches that pass the tests provided by the benchmark.

SWE-bench provides several subsets, and in this study, we used SWE-bench Lite, which contains 300 issue-PR pairs from 11 Python repositories. We refer to these pairs as instances. Among them, 127 come from Django\footnote{https://github.com/django/django} and 80 come from SymPy\footnote{https://github.com/sympy/sympy}. Given its moderate scale and real-world task sources, SWE-bench Lite was suitable for the experiment.

\paragraph{Models}

This study examines whether later LLMs produce patches with better non-functional quality than earlier models from the same model family. Therefore, our model selection was guided by two main factors. The first factor was the release time of the models, which allowed us to compare models from different stages of model evolution. The second factor was model accessibility, which enabled us to compare the performance of both open-source and commercial models.

Based on these criteria, we focused on \emph{Claude} and \emph{DeepSeek}. Claude is a widely used commercial model family, while DeepSeek is a representative open-source model family. For each model family, we selected one earlier model and one later model to support comparison across model iterations. We avoided models that were released too early or showed clearly limited coding capability on SWE-bench. Such models may resolve only a small number of instances on SWE-bench Lite, leaving too few functionally validated patches for non-functional analysis. For this reason, we did not select Claude Opus 3 as the earlier Claude representative. According to the SWE-bench leaderboard\footnote{\url{https://www.swebench.com/}}, Claude Opus 3 achieved a resolved rate of only 11.67\%, which would likely produce a limited sample for subsequent quality analysis. Instead, we used Claude Sonnet 4 as the earlier Claude model.

Following the same selection criteria, the final models were Claude Sonnet 4, Claude Opus 4.6, DeepSeek R1, and DeepSeek V3.2. \autoref{tab:selected-models-lite} summarizes the selected models, their release times, and their reported SWE-bench results. The reported resolved rates in \autoref{tab:selected-models-lite} were taken from the SWE-bench leaderboard and were used only to guide model selection, not as results of our experiment.

\begin{table}[t]
\centering
\caption{Selected models and their reported SWE-bench resolved rates. Retrieved on May 16, 2026.}
\label{tab:selected-models-lite}
\begin{tabular}{lll}
\toprule
\textbf{Model} & \textbf{Release time} & \textbf{Resolved \%}  \\
\midrule
Claude Sonnet 4 & May 2025 & 66.60  \\
Claude Opus 4.6 & Feb. 2026 & 75.60  \\
DeepSeek R1 (0528) & May 2025 & 57.60  \\
DeepSeek V3.2 (high reasoning) & Dec. 2025 & 70.00 \\
\bottomrule
\end{tabular}
\end{table}

\paragraph{Prompts and Configurations}
Our experiments used SWE-agent as the agent framework and adopted the prompt template provided by SWE-bench Lite. Since LLM performance can be sensitive to prompts, using a widely adopted prompt template helps improve the reproducibility and reliability of our experimental results.

In addition, due to budget constraints, we set a maximum cost limit for each instance. To determine an appropriate threshold, we referred to the cost statistics reported on the SWE-bench leaderboard and computed the median cost for the evaluated models. The median per-instance costs were \$0.31 for Claude Opus 4.6 and \$0.36 for DeepSeek V3.2. Based on these values, we set the per-instance cost limit to \$0.50. This threshold gives the models enough budget to complete most tasks while keeping the total evaluation cost manageable. The full prompts and configuration files are available in our replication package \footnote{\url{https://github.com/12xxx21/APSEC_replicate_package.git}}.

\subsection{Patch Quality Assessment}

\paragraph{Functional Validation}

Functional validation serves as the prerequisite for subsequent non-functional quality assessment. Before evaluating the non-functional quality of LLM-generated patches, we first verified whether each generated patch successfully resolved the target issue under the benchmark protocol. We used the official SWE-bench evaluation harness to validate generated patches. The evaluation was conducted in a Docker environment. For each instance, the harness created the repository environment in a container and then attempted to apply the generated patches. Once the generated patch was successfully applied, the harness executed the benchmark tests and output the test results. We considered a generated patch to have passed functional validation only if it could be successfully applied and was marked as resolved by the harness. Patches that failed to apply, failed to resolve the issue, or did not complete the evaluation process were treated as failed and excluded from the non-functional quality evaluation. This step yielded a validated subset, which was used for static analysis and runtime and memory evaluation.

\paragraph{Static Analysis}

After functional validation, we performed static analysis on all patches that passed functional validation using CodeQL\footnote{https://codeql.github.com/} and CodeScene\footnote{https://github.com/empear-analytics} to identify potential quality issues, security risks, and maintainability changes introduced by the generated patches. CodeQL is a widely used semantic code analysis engine that supports query-based static analysis. We used its official Python \textit{security-and-quality} query suite to detect security, reliability, and maintainability issues. CodeScene, in contrast, was used to assess maintainability through its Code Health measure.

Static analysis can be performed at different scopes. Repository-wide analysis scans the entire repository and can capture effects on related files and modules. It also requires more time and resources. Incremental analysis examines the modified files and requires fewer resources. It may miss effects beyond those files. We compared the two scopes on a small pilot sample. The comparison covered analysis cost and reported findings. According to the results, we then used repository-wide analysis for CodeQL and incremental analysis for CodeScene.

For CodeQL, we analyzed each repository before and after patch application and then compared the resulting findings to identify \emph{added}, \emph{removed}, and \emph{unchanged} items. For CodeScene, we used the \texttt{cs delta} command, which directly reports the impact of a patch in terms of code health changes and therefore does not require an additional differencing step.

The analysis was executed independently inside a container. For each instance, we cloned the target repository, created a working directory, and checked out the base commit. We applied the same filtering rules to every instance. We excluded non-Python files, tests, documentation, and other files unrelated to production code. After filtering, we restored the files involved in the patch from the Git index. This prevented pruning from causing patch application failures. We stored the CodeQL comparisons and CodeScene delta results in structured JSON files. These files formed the input to the quantitative analysis.

\paragraph{Runtime and Memory Measurements}

Besides the static analysis, we assessed the dynamic quality of functionally correct patches by measuring CPU time and peak memory consumption during test execution. For each patch that passed functional validation, we ran the harness tests five times under the same container environment and calculated the mean CPU time and the mean of the per-run peak memory measurements. CPU time was obtained by summing user and system time from the cgroup v2 CPU statistics, while peak memory was obtained from the cgroup v2 memory high-water mark. When cgroup measurements were unavailable, we used wall-clock runtime as the CPU fallback and the Docker API memory value as the memory fallback.

\subsection{Data Analysis}
\label{sec:da}

\paragraph{Patch Quality Measures}

We conducted the analysis at the patch level and included only patches marked as resolved by the SWE-bench evaluation harness. This yielded 133 patches for Claude Opus 4.6, 108 patches for Claude Sonnet 4, 133 patches for DeepSeek V3.2, and 79 patches for DeepSeek R1.

For each resolved patch, we derived static indicators from both CodeQL and CodeScene. For CodeQL, we compared the findings before and after patch application and computed the net change as the number of added findings minus the number of removed findings. A positive value means that the patch leaves more findings than the original version, while a negative value indicates that some findings were removed. We also grouped CodeQL findings by their rule tags, including \emph{security}, \emph{maintainability}, \emph{reliability}, and \emph{other}. For CodeScene, we used the \emph{delta} analysis and selected the minimum Code Health change among the files modified by the patch. This value represents the most negative maintainability change found in the modified files. For the performance analysis, we used the mean CPU time and mean peak memory usage from five repeated harness test runs.

We then classified each patch as a regression, an improvement, or a neutral change. For CodeQL, a positive net change indicates a regression, while a negative net change indicates an improvement. For CodeScene, a negative Code Health change indicates a regression, while a positive change indicates an improvement. A zero change, or no reported CodeScene change, was treated as neutral. When analyzing issue patterns, we recorded whether each type of issue appeared in a patch rather than using the total number of findings. This prevents patches with many findings of the same type from having too much influence on the results.

\paragraph{Comparison on Commonly Resolved Instances}

Our primary analysis compared models on the instances resolved by the earlier and later models within each family. This controlled for differences in the tasks included in each model's resolved set. The comparison included 71 instances for Claude and 56 instances for DeepSeek.

For each metric, we calculated the paired difference between the later and earlier models. We used a two-sided signed-rank randomization test and reported the matched rank-biserial correlation. For CPU time and peak memory, we also estimated a 95\% bootstrap confidence interval for the median paired difference. A positive CPU or memory difference indicates greater resource use by the later model. A positive CodeScene difference indicates better Code Health. A positive CodeQL difference indicates more findings. We applied Holm correction to the static tests and the dynamic tests separately.

As an additional sensitivity analysis, we used the 30 instances resolved by all four models. For each instance, we compared the mean value of the two later models with that of the two earlier models. Because this analysis combined the Claude and DeepSeek families, we used it only to check the robustness of the main family-level results.

\paragraph{CodeQL Rules and CodeScene Categories}

To address RQ2, we examined newly reported CodeQL rules and degraded CodeScene categories on the matched sets. For each CodeQL rule and CodeScene category, we recorded whether it appeared at least once in a patch. We then compared the proportion of patches containing each rule or category between the earlier and later models. We used the exact McNemar test based on discordant pairs and applied the Holm correction separately within each tool and model family.

We also examined raw event counts on the full resolved sets. Because the models resolved different sets of instances, these results are mainly used to describe the most common CodeQL rules and CodeScene categories for each model.

\paragraph{Sensitivity Analysis of Resolved Sets and Patch Size}

To examine whether the commonly resolved instances differed from the instances resolved by only one model, we divided the instances in each model family into three groups: instances resolved by both models, instances resolved only by the later model, and instances resolved only by the earlier model.

We compared the three groups using CPU time, peak memory, the number of changed lines in the gold patch, and the number of files modified by the gold patch. We also compared their repository distributions. For numerical measures, we used permutation tests for differences in medians and applied the Holm correction to account for multiple comparisons.

Within the commonly resolved instances, we also compared the generated patches in terms of changed lines, files modified, and patch hunks. In addition, we used exploratory Spearman permutation tests to examine the relationship between patch churn and the number of static analysis findings.

\paragraph{Gold Baseline}

We also applied the static analysis pipeline to the gold patches from SWE-bench Lite. The gold set covers the 194 instances resolved by at least one evaluated model. Each analyzed LLM patch therefore has a gold patch for the same instance. The gold patches serve as a reference and not as a competing model group. They show the distribution of static-analysis findings in gold patches for the same set of instances.

\section{Results}
\subsection{Functional Outcomes}
\autoref{tab:functional_overview} presents the functional correctness of the evaluated models on SWE-bench Lite. \textit{Submitted} denotes the number of instances for which the model generated a patch. \textit{Completed} denotes the number of instances that completed the evaluation process. \textit{Resolved} denotes the number of instances for which the generated patch fixed the issue according to the benchmark test suites.

Overall, Claude Opus 4.6 and DeepSeek V3.2 achieved the highest number of resolved instances, with 133 resolved cases each. Among the earlier models, Claude Sonnet 4 resolved 108 instances, while DeepSeek R1 resolved 79 instances. The models also differed in their completion counts. DeepSeek V3.2 completed 214 evaluations and DeepSeek R1 completed 212. Claude Opus 4.6 completed 178 evaluations and Claude Sonnet 4 completed 174. All models were attempted on the full set of 300 SWE-bench Lite instances, but DeepSeek R1 only produced submissions for 231 instances. This lower submission count was caused by execution failures during environment setup and tool use. Some runs exceeded the cost limit before producing a patch. We computed the resolved rate over 300 instances for every model. These results define the resolved patch sets used in the following non-functional quality analysis.

\begin{table*}[ht]
    \centering
    \caption{The functional outcomes of the models on SWE-bench Lite. The later models are highlighted.}
    \begin{tabular}{lcccc}
    \toprule
      \textbf{Model} & \textbf{Submitted} & \textbf{Completed} & \textbf{Resolved} & \textbf{Resolved Rate (\%)} \\
      \midrule
     \rowcolor{lightgray}Claude Opus 4.6 & 300 & 178 & 133 & 44.33 \\
       Claude Sonnet 4  & 300 & 174 & 108 & 36.00 \\
     \rowcolor{lightgray}  DeepSeek V3.2 & 300 & 214 & 133 & 44.33 \\
       DeepSeek R1 & 231 & 212 & 79 & 26.33 \\
       \rowcolor{yellow!25}Gold patches & 300 & 300 & 300 & 100.00 \\
       \bottomrule
    \end{tabular}

    \label{tab:functional_overview}
\end{table*}

\subsection{Overlap and Composition of the Resolved Sets}
\label{sec:resolved-overlap}

The resolved sets of the models overlapped only partially. Claude Opus 4.6 and Claude Sonnet 4 resolved 71 instances in common. Opus 4.6 resolved an additional 62 instances, while Sonnet 4 resolved an additional 37 instances. DeepSeek V3.2 and DeepSeek R1 resolved 56 instances in common. V3.2 resolved an additional 77 instances, while R1 resolved an additional 23 instances. Only 30 instances were resolved by all four models.

The full resolved sets therefore contained different tasks and repository distributions. For example, 57 of the 71 instances resolved by both Claude models came from Django, whereas 31 of the 62 instances resolved only by Opus 4.6 came from SymPy.

\begin{figure}[t]
    \centering
    \includegraphics[width=\linewidth]{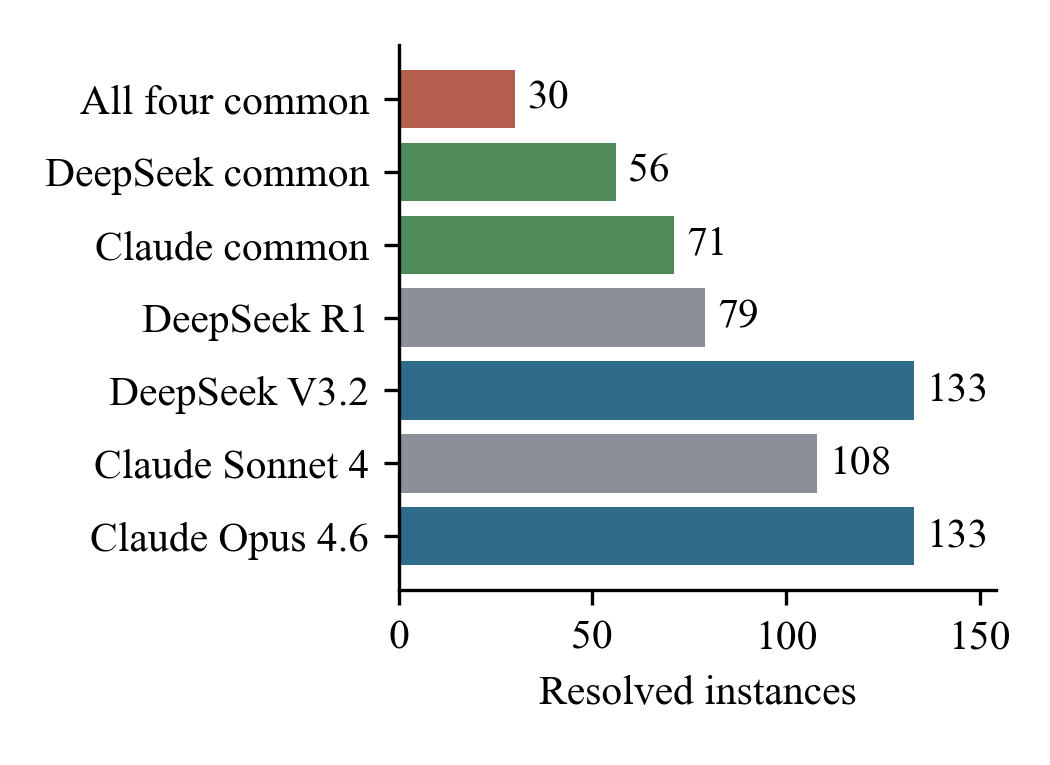}
    \caption{Numbers of instances resolved by each model and included in the common analysis sets. The intersections within the Claude and DeepSeek families contain more instances than the intersection across all four models.}
    \label{fig:resolved-overlap}
\end{figure}

For Claude, the instances resolved only by Opus 4.6 had lower gold-patch peak memory than the instances resolved by both models, with medians of 174.6 and 235.0 MiB, respectively ($p<0.001$ after Holm correction). The instances resolved only by Sonnet 4 had higher gold-patch CPU time than the common instances, with medians of 6.85 and 5.97 seconds, respectively ($p=0.015$ after Holm correction). No other comparison remained significant after correction, including all comparisons for DeepSeek.

\subsection{Static Quality Changes}
\label{sec:direction-share}

\begin{figure*}[t]
    \centering
    \includegraphics[width=0.95\linewidth]{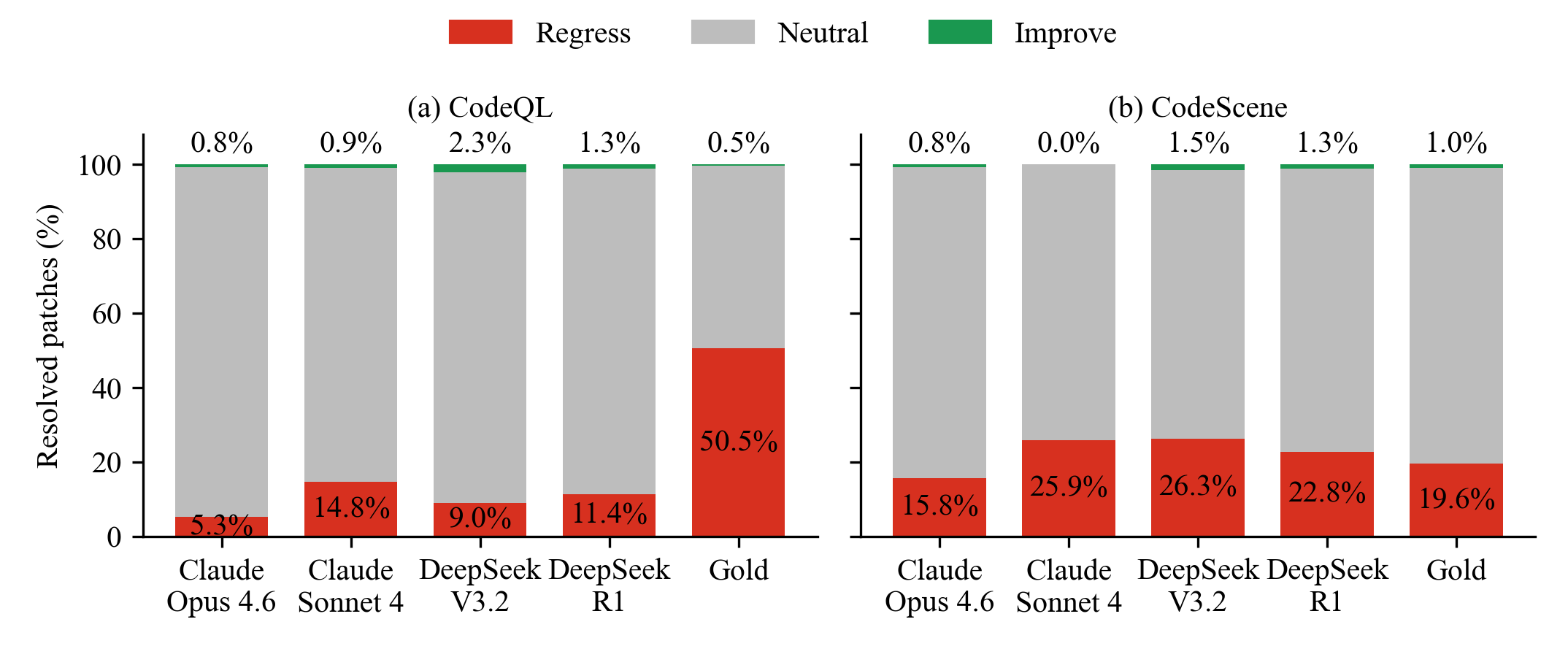}
    \caption{Share of resolved patches that regress, stay neutral, or improve under CodeQL (a) and CodeScene (b). Regression rates are higher than improvement rates for both tools. Gold has a different relative position under the two tools.}
    \label{fig:direction-share}
\end{figure*}

Before comparing later and earlier models, we first examined the quality changes of the resolved patches for each model. Using the change classifications defined in \autoref{sec:da}, \autoref{fig:direction-share} reports the shares of regression cases, neutral cases, and improvement cases for each model and for the gold baseline.

Most resolved patches were neutral under both tools. For the four LLM models, the neutral share ranged from 72\% to 94\%. Improvements were rare and remained below 2.5\% for all models. For CodeScene, we classified each patch by the worst file among the files it touched. A patch was counted as an improvement only when even this worst file improved. 
Regressions were more common than improvements. For the four LLM models, CodeQL reported regressions for 5.3\% to 14.8\% of patches. CodeScene reported higher rates, from 15.8\% to 26.3\%. The gold baseline reversed this pattern. CodeQL classified 50.5\% of gold patches as regressions, while CodeScene classified 19.6\%. This indicated that gold patches often affected CodeQL findings but less often reduced the CodeScene Code Health score.

\subsection{Static Analysis Across Model Generations}
\label{sec:rq1-quality}

\autoref{tab:matched-static} presents the primary comparisons on the instances resolved by both models within each family. The median paired difference was zero for both static measures in both model families. None of the comparisons remained significant after Holm correction. The additional analysis of the 30 instances resolved by all four models produced the same overall result.

Most of the CodeQL pairs had a difference of zero. This included 66 of 71 Claude pairs and 52 of 56 DeepSeek pairs. Only five Claude pairs and four DeepSeek pairs were non-zero. With so few non-zero pairs, the paired test could not reach significance. Even if every non-zero pair pointed in the same direction, the smallest possible two-sided $p$-value would have been about 0.06 for Claude and about 0.13 for DeepSeek. A value of $p=1.000$ therefore reflected the small number of non-zero pairs and did not constitute evidence that the two models were equal. As shown by the number of non-zero pairs reported in \autoref{tab:matched-static}, most matched comparisons show no change in the static analysis results. These results provide no clear evidence that later models improve static quality on commonly resolved tasks, although small differences may still exist.

\begin{table}[t]
\centering
\setlength{\tabcolsep}{4pt}
\caption{Static analysis results for the instances resolved by both models. $\widetilde{\Delta}$ is the median difference between the later and earlier models. $n_{\neq 0}$ is the number of non-zero pairs, and $r$ is the matched rank-biserial correlation. $p_{\min}$ is the smallest two-sided $p$-value the paired test can reach given $n_{\neq 0}$; $p_H$ is the Holm-adjusted value. The all-four CodeQL effect size is based on only one non-zero pair.}
\label{tab:matched-static}
\begin{tabular}{llrrrrrr}
\toprule
\textbf{Comparison} & \textbf{Metric} & \textbf{$n$} & \textbf{$n_{\neq 0}$} & \textbf{$\widetilde{\Delta}$} & \textbf{$r$} & \textbf{$p_{\min}$} & \textbf{$p_H$} \\
\midrule
Claude & CodeQL & 71 & 5 & 0.00 & $-0.07$ & 0.063 & 1.000 \\
Claude & CodeScene & 71 & 10 & 0.00 & 0.64 & 0.002 & 0.504 \\
DeepSeek & CodeQL & 56 & 4 & 0.00 & 0.50 & 0.125 & 1.000 \\
DeepSeek & CodeScene & 56 & 11 & 0.00 & 0.03 & 0.001 & 1.000 \\
All four & CodeQL & 30 & 1 & 0.00 & $-1.00$ & 1.000 & 1.000 \\
All four & CodeScene & 30 & 6 & 0.00 & 0.33 & 0.031 & 1.000 \\
\bottomrule
\end{tabular}
\end{table}

The analysis of the full resolved sets gave the previously observed CodeQL result of $p=0.039$ with a negligible Cliff's $\delta=-0.06$. For CodeScene, it gave $p=0.393$ with $\delta=0.03$. These differences disappeared after matching the instances, suggesting that they were related to task composition rather than model generations.

\subsection{Runtime and Memory Across Model Generations}
\label{sec:rq1-dynamic}

The comparisons on the instances resolved by both models differed from those based on the full resolved sets. \autoref{tab:matched-dynamic} presents the paired results for CPU time and peak memory. For Claude, Opus 4.6 used a median of 0.048 seconds more CPU time than Sonnet 4 on the same 71 instances. It also used 4.524 MiB more peak memory. The median ratios were 1.007 for CPU time and 1.020 for memory. Both differences remained significant after correction. For DeepSeek, the CPU difference was not significant. V3.2 used a median of 0.512 MiB more memory than R1. The ratio was 1.002, and the adjusted $p$-value was 0.002. Excluding the four DeepSeek instances that required fallback measurements left 52 pairs. The median CPU difference was then 0.002 seconds (95\% CI [$-0.068$, 0.025], adjusted $p=1.000$), while the median memory difference was 0.325 MiB (95\% CI [0.132, 0.874], adjusted $p=0.012$). Excluding these instances therefore did not change the interpretation. The CPU time difference was still not significant, while V3.2 continued to show a small increase in peak memory.

On the 30 instances resolved by all four models, CPU time did not differ significantly. However, the mean peak memory of the two later models was 2.876 MiB higher. The ratio was 1.012, and the adjusted $p$-value was below 0.001. The absolute differences were small and did not show a general resource advantage for the later models.

\begin{figure*}[!htbp]
    \centering
    \includegraphics[width=0.95\linewidth]{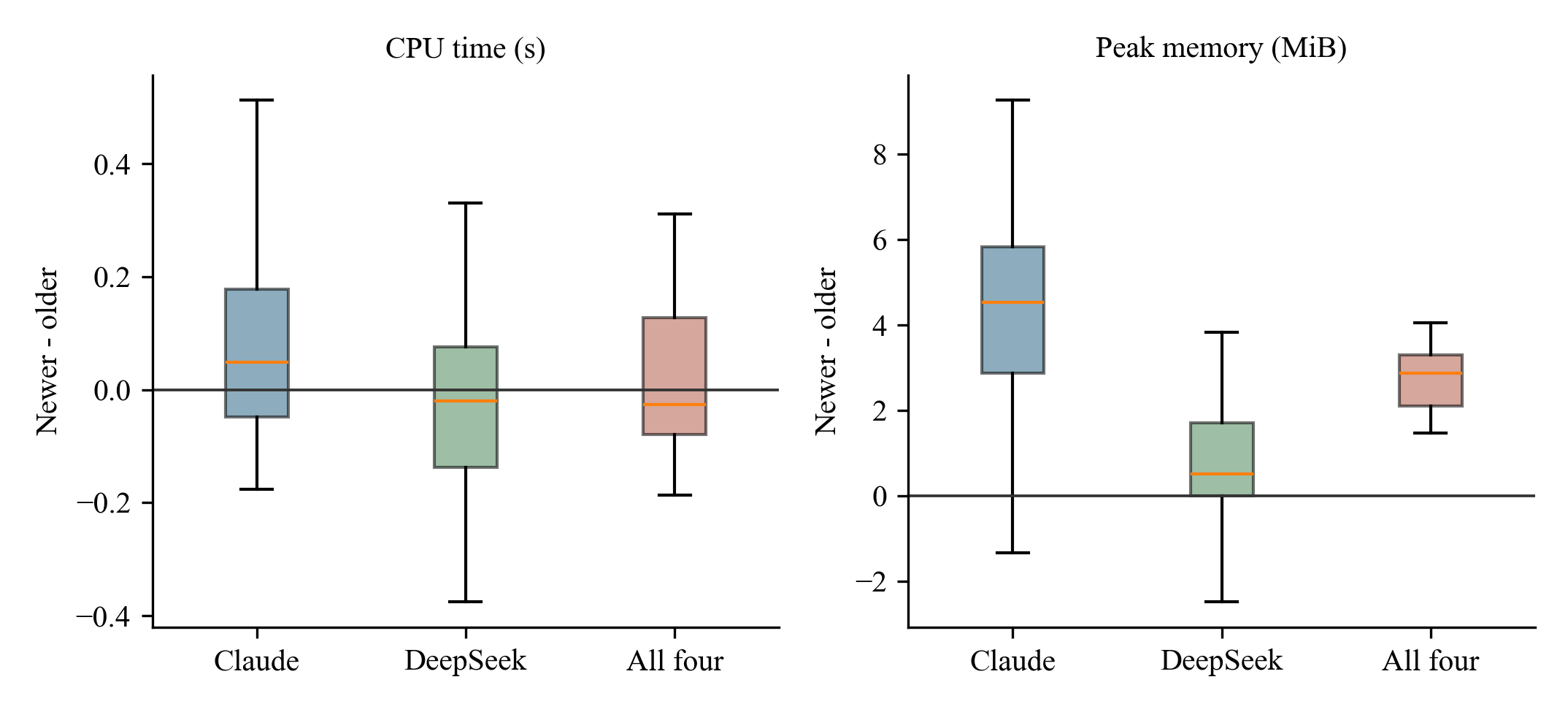}
    \caption{Paired differences in CPU time and peak memory for the instances resolved by both models. Positive values indicate greater resource use by the later model. Extreme observations are omitted from the whiskers for readability, but all observations are included in the statistical tests.}
    \label{fig:dynamic-perf}
\end{figure*}

\begin{table}[!htbp]
    \centering
    \caption{Runtime and memory measurements for each model on its full resolved set. Each cell reports the mean and median, in parentheses, of the per-instance means calculated across five test runs.}
    \label{tab:dynamic-perf}
    \begin{tabular}{lrrr}
    \toprule
    \textbf{Model} & \textbf{n} & \textbf{CPU (s)} & \textbf{Memory (MiB)} \\
    \midrule
    \rowcolor{lightgray}Claude Opus 4.6 & 133 & 10.18\,(5.93) & 215.4\,(218.0) \\
    Claude Sonnet 4                      & 108 & 17.87\,(5.97) & 284.9\,(242.6) \\
    \rowcolor{lightgray}DeepSeek V3.2   & 133 &  9.05\,(5.92) & 206.2\,(214.6) \\
    DeepSeek R1                          &  79 & 16.47\,(6.30) & 241.3\,(215.8) \\
    \rowcolor{yellow!25}Gold patches & 194 & 15.06\,(6.01) & 240.8\,(215.9) \\
    \bottomrule
    \end{tabular}
\end{table}

\begin{table*}[t]
\centering
\caption{Runtime and memory results for the instances resolved by both models. Differences compare the later model with the earlier model. $\widetilde{\Delta}$ is the median paired difference, $r$ is the matched rank-biserial correlation, and the CI is the 95\% bootstrap confidence interval for the median paired difference. $p_H$ is the Holm-adjusted $p$-value.}
\label{tab:matched-dynamic}
\begin{tabular}{llrrrcr}
\toprule
\textbf{Comparison} & \textbf{Metric} & \textbf{$n$} & \textbf{$\widetilde{\Delta}$} & \textbf{$r$} & \textbf{95\% CI} & \textbf{$p_H$} \\
\midrule
Claude & CPU (s) & 71 & 0.048 & 0.42 & [0.008, 0.100] & 0.005 \\
Claude & Memory (MiB) & 71 & 4.524 & 0.93 & [3.714, 5.642] & $<.001$ \\
DeepSeek & CPU (s) & 56 & $-0.020$ & $-0.22$ & [$-0.076$, 0.020] & 0.329 \\
DeepSeek & Memory (MiB) & 56 & 0.512 & 0.53 & [0.162, 1.099] & 0.002 \\
All four & CPU (s) & 30 & $-0.026$ & 0.06 & [$-0.053$, 0.075] & 0.778 \\
All four & Memory (MiB) & 30 & 2.876 & 0.99 & [2.350, 3.201] & $<.001$ \\
\bottomrule
\end{tabular}
\end{table*}

\subsection{CodeScene Categories by Model}
\label{sec:per-model-cs-categories}

On the instances resolved by both Claude models, \textit{Lines of Code in a Single File} appeared in 19.7\% of the Opus 4.6 patches and 25.4\% of the Sonnet 4 patches. \textit{Number of Functions in a Single Module} had the same difference, while \textit{Bumpy Road Ahead} decreased by $4.2$ percentage points.

On the instances resolved by both DeepSeek models, the largest difference was $+3.6$ percentage points for \textit{Lines of Code in a Single File}. \textit{Complex Method} appeared in 21.4\% of the V3.2 patches and 19.6\% of the R1 patches, corresponding to a difference of $+1.8$ percentage points.

None of the differences in CodeScene category prevalence remained significant after Holm correction. 

\subsection{CodeQL Rules by Model}
\label{sec:per-model-signatures}

The patch prevalence analysis assigned one binary observation to each patch for each CodeQL rule. For Claude, the largest differences favored Opus for \texttt{missing-equals} (8.5\% vs. 14.1\%) and \texttt{mixed-returns} (12.7\% vs. 16.9\%). For DeepSeek, \texttt{conflicting-attributes} occurred in 25.0\% of V3.2 patches and 16.1\% of R1 patches. The differences for \texttt{cyclic-import} and \texttt{unexpected-raise-in-special-method} were both 3.6 percentage points. None of these differences remained significant after Holm correction.

In the full resolved sets, raw event counts still placed \texttt{cyclic-import} and \texttt{conflicting-attributes} among the most frequent rules. However, the same rule often appeared in both added and removed findings. This can happen when code movement changes the location of a finding. Patch prevalence and matched comparisons reduce this measurement limitation, but they do not remove it completely.

\begin{figure*}[t]
    \centering
    \includegraphics[width=0.9\linewidth]{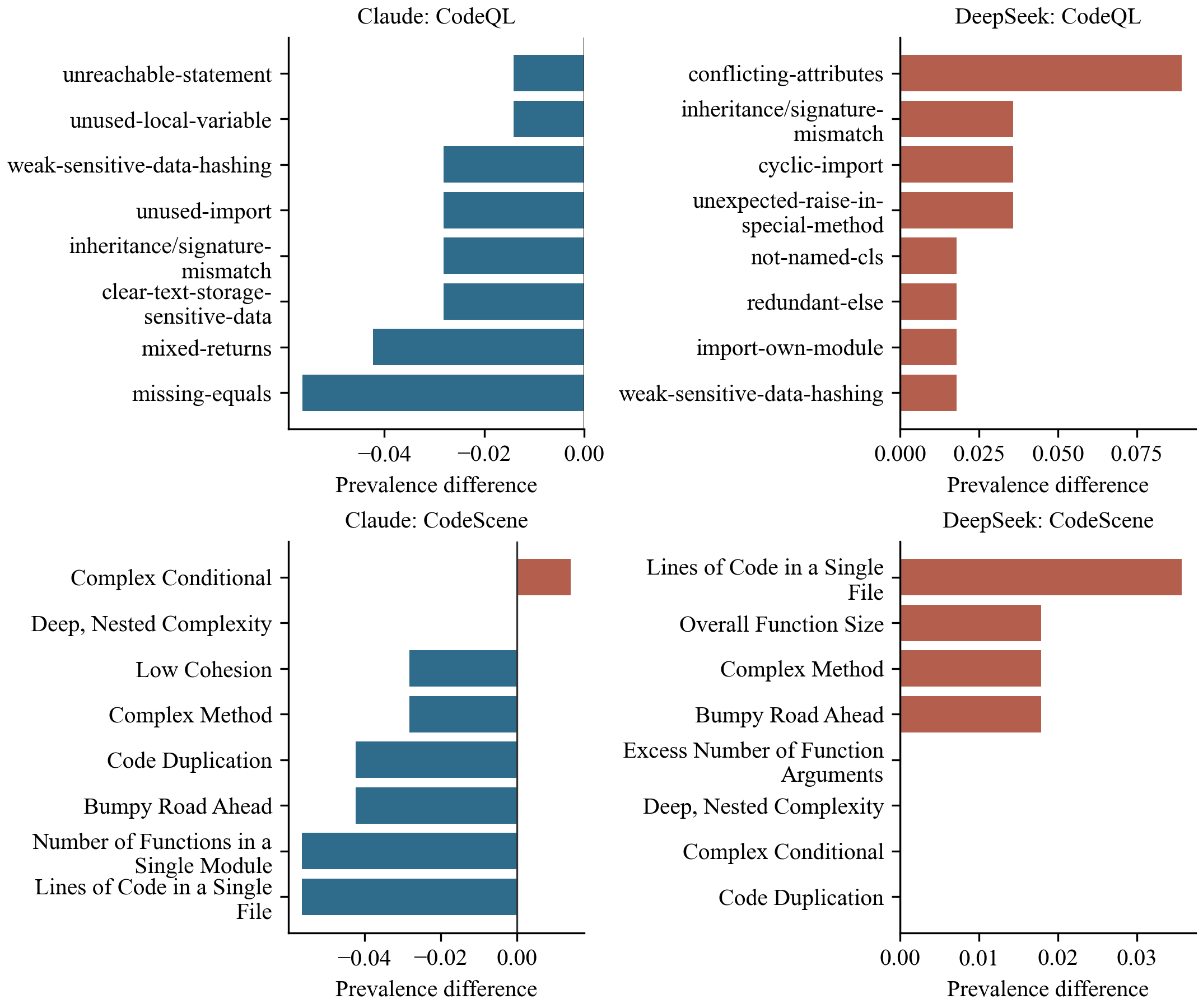}
    \caption{Largest differences in patch prevalence for CodeQL rules and CodeScene categories. Positive values indicate higher prevalence for the later model. None of the differences remains significant after Holm correction.}
    \label{fig:matched-patterns}
\end{figure*}

\subsection{Sensitivity Analysis of Patch Sizes}
\label{sec:patch-size}

Patch size showed different patterns across the two model families. On the Claude common set, Opus 4.6 changed fewer lines than Sonnet 4, with medians of three and six lines, respectively ($p<0.001$ after Holm correction). Opus 4.6 also modified fewer files and used fewer hunks. For DeepSeek, V3.2 changed more lines than R1, with medians of 6.5 and four lines, respectively ($p=0.003$), and also used more hunks.

Patch churn was positively associated with CodeScene degradations for Sonnet 4, DeepSeek R1, and DeepSeek V3.2, with $\rho$ values of 0.37, 0.48, and 0.30, respectively. It was also associated with added CodeQL findings for V3.2 ($\rho=0.31$). All associations remained significant after Holm correction.

\section{Discussion}

Comparing the models on the commonly resolved instances gives a different view from the results based on the full resolved sets. Although the later models resolve more instances, we find no clear evidence that they produce better CodeQL or CodeScene results on commonly resolved tasks. The median paired differences for all static measures are zero, and the significant CodeQL result from the full resolved sets disappears when the same instances are compared. Resolved rate and static quality measures therefore capture different aspects of model performance. However, this analysis does not cover instances resolved only by the later models, so it cannot determine the non-functional quality of the additional patches responsible for the higher resolved rate.

Task composition also affects the resource results. The full resolved sets suggest lower CPU time and peak memory for the later models, whereas the paired comparison shows no consistent CPU advantage and slightly higher peak memory. Under the workloads covered by the SWE-bench Lite tests, these differences are small, with median ratios generally below 1.02.

The CodeQL rule and CodeScene category results show no consistent pattern across the two model families. Some findings are less common in the later Claude patches, while others are more common in the later DeepSeek patches. None remains significant after Holm correction. Patch size may partly contribute to these differences, as the later Claude model produces smaller patches and the later DeepSeek model produces larger patches. Patch churn is also associated with some static analysis findings, although this does not establish a causal relationship.

The differences between CodeQL and CodeScene further show that no single static analysis tool provides a complete view of patch quality. Among the four LLMs, CodeScene classifies a larger proportion of patches as regressions than CodeQL. The gold patches show the opposite pattern, with a higher regression proportion under CodeQL than under CodeScene. This is not unexpected because the two tools measure different properties. CodeQL reports rule-based findings related to security, reliability, maintainability, and other code properties. CodeScene measures structural maintainability through changes in Code Health. A patch may change the presence or reported location of a CodeQL finding without producing a substantial change in the structure measured by CodeScene. Conversely, a patch may reduce the Code Health score without introducing a CodeQL finding. 

Most resolved patches are classified as neutral by both static analysis tools. For CodeQL, frequently reported rules including \texttt{cyclic-import}, \texttt{conflicting-attributes}, \texttt{mixed-returns}, and \texttt{empty-except}, appear among both added and removed findings. In some cases, a patch removes a finding from one location but introduces the same type of finding elsewhere. Such changes modify the code without reducing the overall number of detected issues, and are classified as neutral. Added and removed CodeQL findings therefore do not always correspond directly to the introduction or removal of an underlying issue. This also contributes to the large number of patches with a net change of zero.

\section{Threats to Validity}
\paragraph{Construct Validity}

This study uses CodeQL, CodeScene, CPU time, and peak memory usage as proxies for non-functional quality. These measurements capture important aspects of code quality, but they do not provide a complete assessment. CodeQL reports rule-based findings, while CodeScene reports changes in Code Health. Both tools depend on their own definitions. CPU and memory cover only the paths exercised by the SWE-bench tests. They are not general performance benchmarks. The memory differences are statistically significant, but their absolute values are small. Their magnitude is close to the run-to-run variation of the measurement. Their practical significance is therefore uncertain. The functional prompt measures the quality produced during bug fixing. It does not measure a model's ability to optimize quality when asked.

The analysis of CodeQL finding changes also depends on how findings are matched before and after patch application. When a patch moves or rewrites code, a finding at the original location may be recorded as removed, while the same rule is reported again at the new location. This can affect the added and removed finding counts, especially for rules that are sensitive to code location or local context. We therefore treat changes in CodeQL findings as changes in the static analysis report, rather than direct evidence of semantic quality improvement or degradation.

\paragraph{Internal Validity}

The selection of resolved instances is the main internal threat. Matching within each family ensures that the compared patches address the same instances. The analysis of the 30 common instances provides an additional check. However, these analyses focus on tasks that both models can solve. They do not show the quality of patches for tasks solved by only one model. The selection analysis also shows clear repository differences between the common and exclusive sets. We therefore report the results from the full resolved sets only as descriptive evidence.

The model pairs do not differ only in release time. Sonnet 4 and Opus 4.6 belong to different Claude tiers. R1 and V3.2 differ in their model design and reasoning configuration. The observed results may therefore include effects from the model tier, architecture, or interaction with the agent. The \$0.50 cost limit may also affect the models differently because their prices and reasoning lengths differ. Its influence cannot be ruled out without an additional evaluation using a higher limit.

All patches were evaluated with the same harness and container setup. Runtime and memory can still be affected by system noise. We therefore measured each patch five times. The median coefficients of variation range from 1.1\% to 1.3\% for CPU and from 0.4\% to 3.2\% for memory. The indicators are skewed and contain many zero values. This leaves few non-zero static pairs and limits the statistical power. We therefore used paired randomization tests, bootstrap intervals, patch prevalence, and Holm correction. Patch churn is associated with some findings. However, this association differs across families and cannot be fully separated from model behavior.

\paragraph{External Validity}

The study is based on SWE-bench Lite, which contains real bug-fixing tasks from Python repositories. This benchmark is suitable for studying repository-level patch generation, but it does not cover all software engineering tasks. In particular, the results may not generalize to large refactorings, feature development, security-sensitive changes, or projects written in other programming languages.

The security-related results also reflect the task distribution of SWE-bench Lite. Few security rules appear among the most frequent CodeQL findings in our results. This does not imply that LLM-generated patches are generally safe from security risks. Rather, it reflects the types of tasks and code changes present in SWE-bench Lite.

The evaluated set is limited to four models from two provider families. It supports two case studies rather than a general claim about all LLM generations. SWE-bench Lite also contains many instances from Django and SymPy. The matched subsets have different repository compositions. Different languages, benchmarks, agent frameworks, decoding settings, or tool-use mechanisms may produce different patch characteristics.

Finally, this study evaluates patches generated under a prompt focused on functional bug repair. The models are not explicitly instructed to improve security, maintainability, or performance. A quality-focused prompt, an additional review step, or an iterative refinement process could produce different results. The conclusions therefore characterize the repair setting, agent configuration, and benchmark workload used in this study. They do not represent the full ability of the models to improve non-functional software quality.

\section{Conclusion}

This study compared the non-functional characteristics of functionally correct patches produced by earlier and later Claude and DeepSeek models. All patches were evaluated under the same SWE-agent repair setting and with the same static analysis and resource measurement pipelines. The primary analysis used 71 instances resolved by both Claude models and 56 instances resolved by both DeepSeek models, with an additional sensitivity analysis on 30 instances resolved by all four models.

The later models showed no consistent improvement in the measured non-functional indicators on commonly resolved tasks. Static differences were not significant after correction, CPU time differences were small and inconsistent, and peak memory was slightly higher for the later models under the benchmark workload. CodeQL rules, CodeScene categories, and patch size also showed different patterns across the two model families. These results indicate that a higher resolved rate does not necessarily correspond to better non-functional characteristics and that resolved rate, static analysis findings, resource use, and patch characteristics should be reported separately in repository-level evaluations.

Future work should evaluate the instances resolved only by later models to determine whether the additional functional coverage is associated with differences in non-functional quality. It should also include more model families, repositories, and programming languages. Further studies could examine whether quality-focused prompts, static-analysis feedback, or iterative repair and review steps improve the non-functional quality of generated patches.

\balance
\bibliographystyle{IEEEtran}
\bibliography{reference}

\end{document}